\documentclass[modern]{aastex63}

\shorttitle{Discovery and timing of three MSPs}
\shortauthors{Bhattacharyya et al.}

\begin{document}

\title{Discovery and timing of three millisecond pulsars in radio and $\gamma$-rays with the GMRT and \emph{Fermi}-LAT}
\author{B.~Bhattacharyya}
\affil{National Centre for Radio Astrophysics, Tata Institute of Fundamental Research, Pune 411 007, India}
\author{J.~Roy}
\affil{National Centre for Radio Astrophysics, Tata Institute of Fundamental Research, Pune 411 007, India}
\author{T.~J.~Johnson}
\affil{George Mason University, resident at U.S. Naval Research Laboratory}
\author{P.~S.~Ray}
\affil{U.S. Naval Research Laboratory, Washington, DC 20375, USA}
\author{P.~C.~C.~Freire}
\affil{Max-Planck-Institut für Radioastronomie, Bonn, D-53121, Germany}
\author{Y.~Gupta}
\affil{National Centre for Radio Astrophysics, Tata Institute of Fundamental Research, Pune 411 007, India}
\author{D.~Bhattacharya}
\affil{Inter-University Centre for Astronomy and Astrophysics, Pune 411 007, India}
\author{A.~Kaninghat}
\affil{National Centre for Radio Astrophysics, Tata Institute of Fundamental Research, Pune 411 007, India}
\author{B.~W.~Stappers}
\affil{Jodrell Bank Centre for Astrophysics, School of Physics and Astronomy, The University of Manchester, Manchester M13 9PL, UK}
\author{E.~C.~Ferrara}
\affil{UMD and NASA/GSFC}
\author{S.~Sengupta}
\affil{Max-Planck-Institut für Radioastronomie, Bonn, D-53121, Germany}
\affil{Indian Institute of Technology, Kharagpur, West Bengal 721302}
\author{R.~S.~Rathour}
\affil{National Centre for Radio Astrophysics, Tata Institute of Fundamental Research, Pune 411 007, India}
\affil{Nicolaus Copernicus Astronomical Centre, Polish Academy of Sciences, Bartycka 18, PL-00-716 Warsaw, Poland}
\author{M.~Kerr}
\affil{U.S. Naval Research Laboratory, Washington, DC 20375, USA}
\author{D.~A.~Smith}
\affil{Centre d'\'Etudes Nucl\'eaires de Bordeaux Gradignan, IN2P3/CNRS, Universit\'e Bordeaux, BP120, 33175 Gradignan, France}
\author{P.~M.~Saz~Parkinson}
\affil{Santa Cruz Institute for Particle Physics, Department of Physics and Department of Astronomy and Astrophysics, University of California at Santa Cruz, Santa Cruz, CA 95064, USA}
\affil{Department of Physics, The University of Hong Kong, Pokfulam Road, Hong Kong, China}
\affil{Laboratory for Space Research, The University of Hong Kong, Hong Kong, China}
\author{S.~M.~Ransom}
\affil{National Radio Astronomy Observatory, 1003 Lopezville Road, Socorro, NM 87801, USA}
\author{P.~F.~Michelson}
\affil{W. W. Hansen Experimental Physics Laboratory, Kavli Institute for Particle Astrophysics and Cosmology, Department of Physics and SLAC National Accelerator Laboratory, Stanford University, Stanford, CA 94305, USA}

\begin{abstract}
We performed deep observations to search for radio pulsations in the directions of 375 unassociated \emph{Fermi} Large Area Telescope (LAT) $\gamma-$ray sources using the Giant Metrewave Radio Telescope (GMRT) at 322 and 607 MHz. In this paper we report the discovery of three millisecond pulsars (MSPs), PSR J0248+4230, PSR J1207$-$5050 and PSR J1536$-$4948. We conducted follow up timing observations for $\sim$ 5 years with the GMRT and derived phase coherent timing models for these MSPs. PSR J0248$+$4230 and J1207$-$5050 are isolated MSPs having periodicities of 2.60 ms and 4.84 ms.  PSR J1536$-$4948 is a 3.07 ms pulsar in a binary system with orbital
period of $\sim$ 62 days about a companion of minimum mass 0.32 M$_\sun$. We also present multi-frequency pulse profiles of these MSPs from the GMRT observations. PSR J1536$-$4948 is an MSP with an extremely wide pulse profile having multiple components. 
Using the radio timing ephemeris we subsequently detected $\gamma$-ray pulsations from these three MSPs,
confirming them as the sources powering the $\gamma$-ray emission. For PSR J1536$-$4948 we performed combined radio$-\gamma$-ray timing using $\sim$ 11.6 years of $\gamma$-ray pulse times  of  arrivals (TOAs) along with the radio TOAs. PSR J1536$-$4948 also shows evidence for pulsed $\gamma$-ray emission out to above 25 GeV, confirming earlier associations of this MSP with a $\geq$10 GeV point source.  The multi-wavelength pulse profiles of all three MSPs offer challenges to models of radio and $\gamma$-ray emission in pulsar magnetospheres.
\end{abstract}

\section{Introduction}
\label{sec:intro}

The  Large Area Telescope \citep[LAT,][]{atwood09}, the primary instrument on the \emph{Fermi Gamma-ray Space Telescope},  has been surveying the GeV $\gamma$-ray sky since its scientific activation on August 4, 2008. This has dramatically increased the number of known $\gamma$ ray sources and with each catalog release there have been an increasing number of sources unassociated with any known counterpart likely to be powering the $\gamma$-ray emission. Particularly at high Galactic latitude, many of these sources have proven to be hitherto unknown millisecond pulsars (MSPs) \citep{ray12}. Searching for pulsations of unknown MSPs in the $\gamma$-ray band is extraordinarily computationally expensive, particularly in the case of binaries. While it has proven possible in a few cases (e.g., \citealt{nieder20}, who used astrometric and orbital data provided by optical observations to greatly reduce the necessary number of trials), it is generally far more efficient to first search for radio pulsars in the direction of these sources. Targeted searches for radio pulsations at the position of unassociated LAT point sources coordinated by the \emph{Fermi} Pulsar Search Consortium \citep[PSC,][]{ray12} have resulted in the discovery of 95 radio MSPs so far, including the ones reported here. Finding pulsars powering these sources is important for identifying the nature of the $\gamma$-ray sources and for the astrophysics made possible by timing the newly-discovered pulsars. An identification also rules out more exotic possible sources such as dark matter subhalos \citep{Coronado-Blazquez19}.

Using the LAT sources to guide searches is a powerful technique.
It allows deep searches through long observations as well as allowing multiple visits per source. This is valuable because a pulsar can be missed in a single observation due to scintillation, eclipses, or acceleration in a binary system. The Giant Metrewave Radio Telescope (GMRT\footnote{http://gmrt.ncra.tifr.res.in}) $-$ a multi-element aperture synthesis telescope consisting of 30 antennas each of 45 m diameter, having maximum baseline length of 25 km  \citep{swarup97} $-$ is particularly well suited to this task. 
The low frequency observing capabilities (300--600 MHz) of the GMRT are optimal for sensitive detection of MSPs having steep spectra and typically low values of the dispersion measures. Its design, featuring a large array of small telescopes, provides multiple advantages: (1) wide field of view with incoherent beam (FWHM of 80\arcmin~at 322 MHz, and 40\arcmin~at 607 MHz; ideal for pulsar search observations), (2) high sensitivity coherent beam (4 to 5 times incoherent array beam; good for follow up timing observations) and (3) rapid precise localization ($\sim$ 10\arcsec ) using the imaging capability, even on search observations. 
The semi-major axis of the 95\% confidence error ellipses of the \emph{Fermi}-LAT sources are about $\pm$10\arcmin~(although the exact value is a function of location and integration time). Hence the larger beam width of the GMRT at lower frequencies is of considerable help. This wide beam allows the GMRT to search in a single observation faint LAT sources that are not well localized, something that cannot be done with large single dish telescopes. The prospect of the GMRT in pulsar searches has been demonstrated by the discovery of 30 pulsars in targeted and blind searches at an encouraging pulsar-per-square degree discovery rate \citep[e.g.,][]{ray12,bh13,bh16,bh19}. 
 
In this paper we present the GMRT discoveries, follow-up timing and subsequent discovery of $\gamma$-ray pulsations for three MSPs which are associated with \emph{Fermi}-LAT sources. Section \ref{sec:sources} details the target selection and provides limiting flux densities for the sources from which pulsations were not detected. A list of all the GMRT pointings and corresponding detection limits are presented in the appendix. Section \ref{sec:obs} of this paper details the search and timing observations with the GMRT. Section \ref{sec:discovery} details the discoveries. Section \ref{sec:localisations} presents the more accurate position estimates of these three MSPs by localizing them in the image plane with the GMRT interferometric array. Results from follow up timing studies of the discovered pulsars are reported in Section \ref{sec:timing}. Section \ref{sec:gammaray} presents the results 
from $\gamma$-ray analysis of these pulsars. Section \ref {sec:Discussion} presents the discussion and Section~\ref{sec:Summary} a summary.
A list of all the GMRT pointings and corresponding detection limits are presented in the Appendix.

\section{Source Selection}
\label{sec:sources}

As part of a broader effort coordinated by the \textit{Fermi} PSC, we selected sources from early versions of the \textit{Fermi}-LAT catalogs
(\citealt{abdo10,nolan12} analysis that were not associated with likely $\gamma$-ray emitting counterparts and were visible from the GMRT (see the Appendix for details). 

Using the GMRT, we have performed a targeted radio pulsar survey of 375 unassociated $\gamma$-ray point sources detected by the \emph{Fermi}-LAT. The survey was conducted with observations at 322 MHz and 607 MHz. We aimed to observe the relatively high latitude pointings at 322 MHz as the dispersion broadening and scattering contributions are comparatively lower for the target sky. For this we selected the sources available at a given time span either at 322 or at 607 MHz. In general, we observed each target once with the GMRT either at 322 MHz or at 607 MHz. However, in case of marginal detection of possible millisecond pulsations we conduct confirmation observations. In this effort, we have discovered four MSPs associated with the \emph{Fermi}-LAT $\gamma$-ray sources.  One MSP, the black widow PSR J1544$+$4937, is the first Galactic MSP discovered by the GMRT and has been published elsewhere \citep{bh13}. 
In Section \ref{sec:discovery} of this paper we present discovery details of the remaining three MSPs. We have also independently detected MSP J1446$-$4701 with the Fermi directed searches with the GMRT, which was already discovered in the HTRU survey \citep{keith11}.
In addition, we discovered three in-beam millisecond pulsars, PSR J1120$-$3618, J1646$-$2142 and J1828+0625, which are not associated with the target \emph{Fermi}-LAT $\gamma$-ray sources. The distances of these pulsars from the pointing centres of the \emph{Fermi}-LAT sources are 57\arcmin, 10\arcmin ~and 26\arcmin ~respectively \citep[Refer to Table 2 of][]{roy13}. These serendipitous discoveries will be reported in a follow up paper (Bhattacharyya et al.~\textit{in preparation}).
The details of our radio observations for all 375 \emph{Fermi}-LAT sources
and corresponding 10$\sigma$ detection limit for each source are also presented in the Appendix.
\section{Observations and Pulsation Search Analysis}
\label{sec:obs}

\begin{figure}
\begin{center}
\includegraphics[width=6.5in,angle=0]{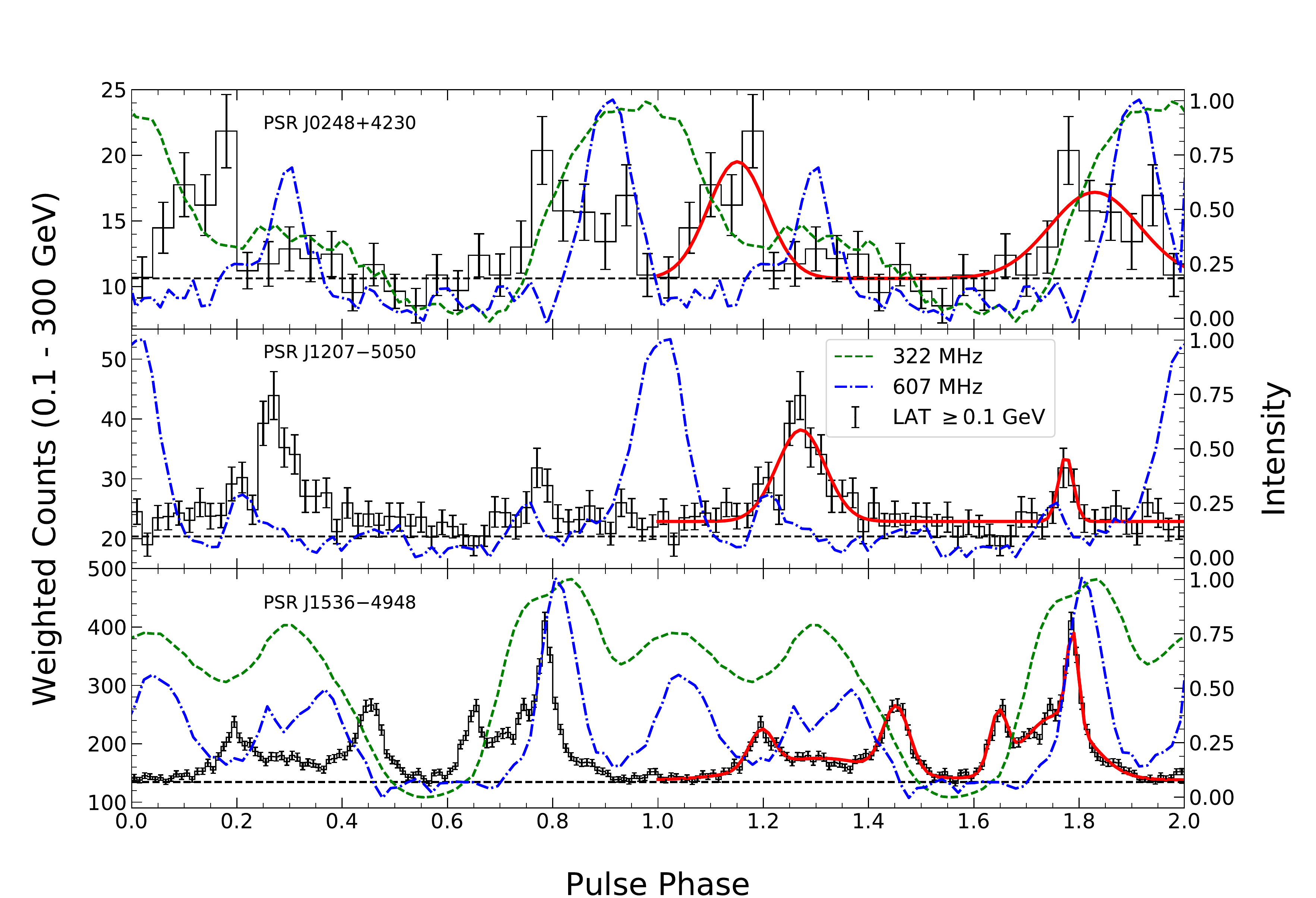}
\caption{Radio and $\gamma$-ray pulse profiles for PSRs J0248+4230 (top), J1207$-$5050 (middle), and J1536$-$4948 (bottom).  The black histograms (left y-axis) show the weighted LAT counts from 0.1 to 300 GeV within 3$^{\circ}$ of the respective pulsar position with the black dashed horizontal line giving the estimated background level \citep[calculated as in][]{2PC}. The solid red lines on the right half of each plot show the results of the pulse profile fits described in the text.  The green dashed curves show the 322 MHz radio profiles, when available, and the blue dash-dot curves show the 607 MHz profile (both for highest signal-to-noise detection and using the right y-axis giving relative intensity in arbitrary units).  The data spanning pulse phases 0 to 1 is duplicated over pulse phases from 1 to 2, to more easily show features occurring near a pulse phase of 1.}
\label{fig:multiprofiles}
\end{center}
\end{figure}

\begin{figure}
\begin{center}
\includegraphics[width=5in,angle=-90]{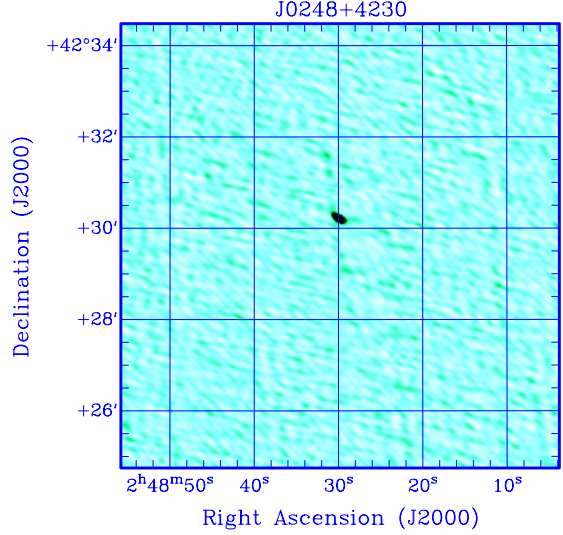}
\caption{The on-off gating image of PSR J0248+4230 using the coherently dedispersed MSP gating correlator. The MSP is localised with $\pm$ 8\arcsec accuracy at 13$\sigma$ detection significance.}
\label{fig:J0248_img}
\end{center}
\end{figure}
\begin{figure}
\begin{center}
\includegraphics[width=5in,angle=-90]{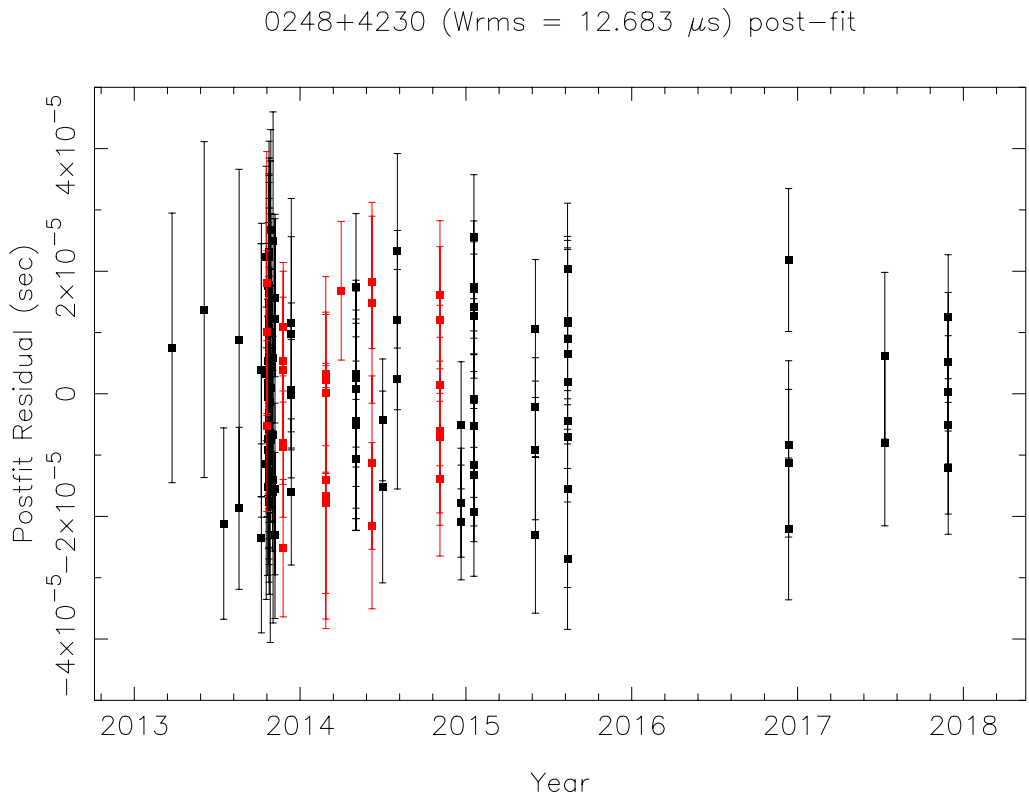}
\caption{Radio timing residuals for PSR J0248$+$4230 from the GMRT observations at 322 MHz (black points) and 607 MHz (red points) with bandwidth of 32 MHz using the GMRT legacy system.}
\label{fig:residual_J0248}
\end{center}
\end{figure}
\begin{figure}
\begin{center}
\includegraphics[width=5in,angle=-90]{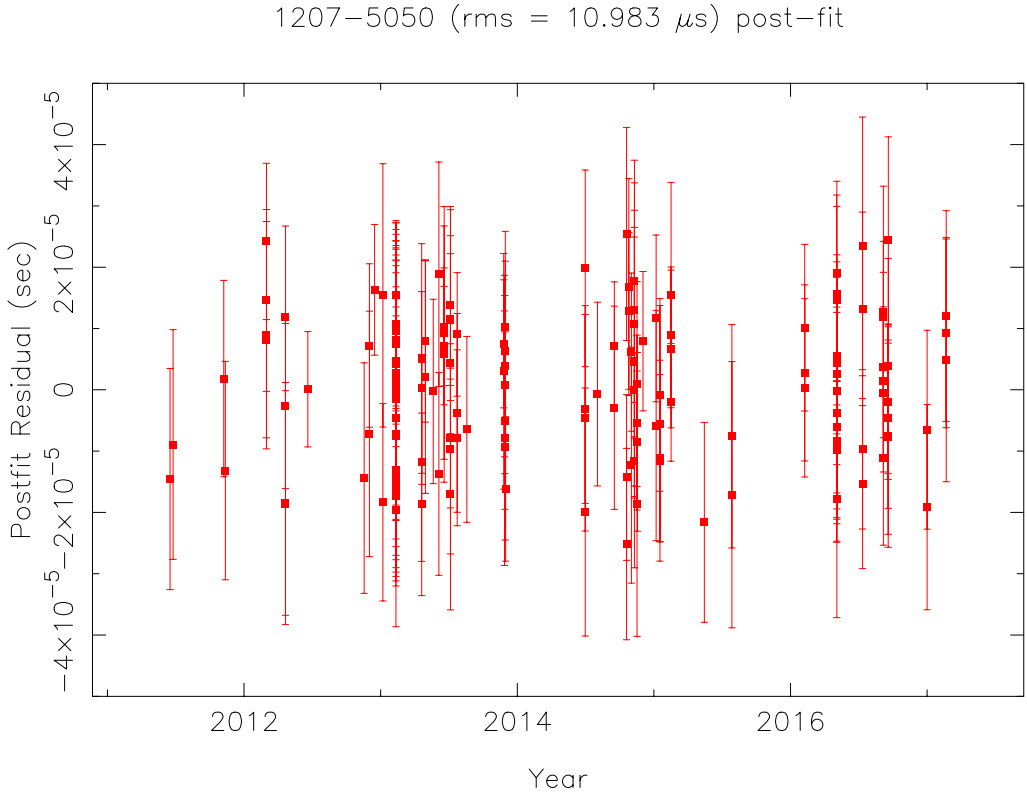}
\caption{Radio timing residuals for PSR J1207$-$5050 from the GMRT observations at 607 MHz (red points) with a 32 MHz bandwidth using the GMRT legacy system. PSR J1207$-$5050 was not detected at 322 MHz in any of our observations with the GMRT.}
\label{fig:residual_J1207}
\end{center}
\end{figure}
\begin{figure}
\begin{center}
\includegraphics[width=5in,angle=-90]{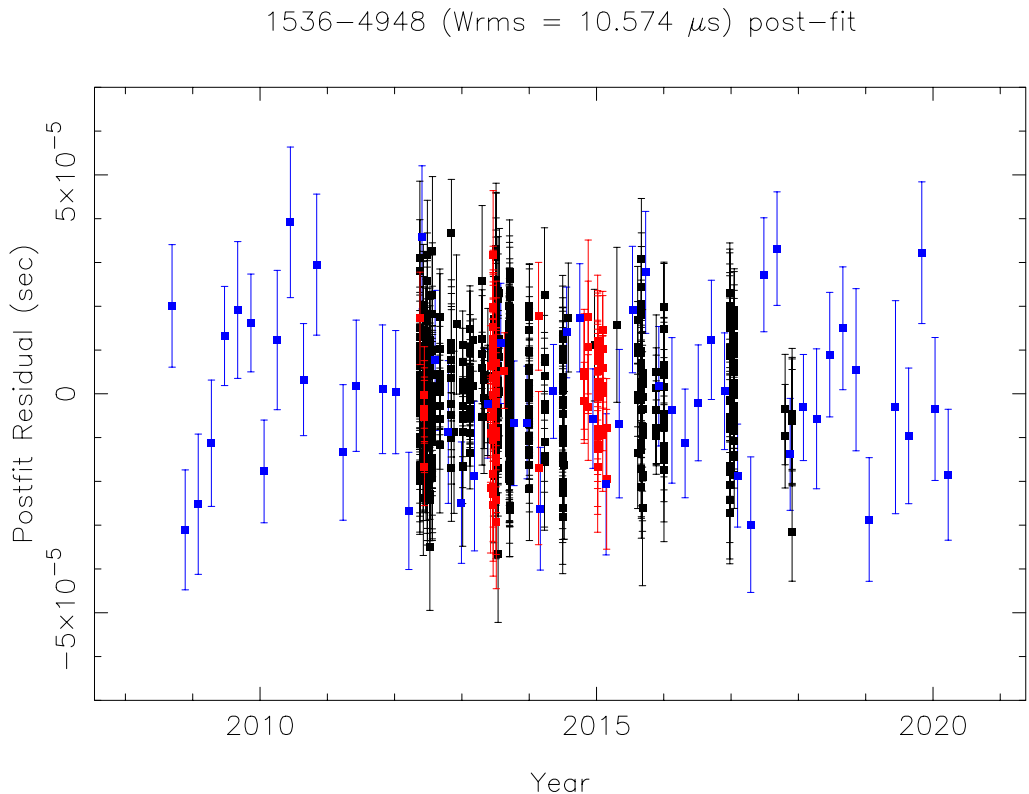}
\caption{Timing residuals for PSR J1536$-$4948 from the GMRT observations at 322 MHz (black points) and 607 MHz (red points) and LAT observations from 0.1 to 300 GeV (blue points).}
\label{fig:residual_J1536}
\end{center}
\end{figure}

\begin{deluxetable}{lccccccccccccccc}
\caption{Parameters of pulsars discovered in \textit{Fermi} directed survey with the GMRT\label{discovery}}
\tablehead{\colhead{Pulsar}            & \colhead{$P$}       & \colhead{$\dot{P}$}                & \colhead{DM}           & \colhead{$S_{322}$\tablenotemark{$\dagger$}}   & \colhead{$S_{607}$\tablenotemark{$\ddagger$}} & \colhead{spectral index} \tablenotemark{$\alpha$} \\
\colhead{} & \colhead{(ms)}   & \colhead{(s/s)} & \colhead{(pc cm$^{-3}$)}   & \colhead{(mJy)}                  & \colhead{(mJy)}                   
}
\startdata
PSR J0248$+$4230  & 2.600  & 1.68$\times$10$^{-20}$  & 48.2636(1)       & 7.5                    & 0.9                   &  $-$3.34(6) \\
PSR J1207$-$5050  & 4.842  & 6.06$\times$10$^{-21}$ & 50.5972(6)       & $<$0.38$^{\ast}$                    & 0.5                    &  $>$0.43(6) \\
PSR J1536$-$4948  & 3.079  & 2.12$\times$10$^{-20}$ & 38.0016(7)       & 24.7                   & 4.0                    &  $-$2.86(6) \\
\enddata
\tablenotetext{\dagger}{Flux density at 322 MHz without primary beam correction.}
\tablenotetext{\ddagger}{Flux density at 607 MHz without primary beam correction.}
\tablenotetext{\ast}{10$\sigma$ non-detection limit at 322 MHz for 30-minutes of GMRT observations using 17 antennas in phased array}
\tablenotetext{\alpha}{ The numbers in the parenthesis are uncertainties in preceding digit. Error on spectral index is calculated considering a typical 10\% uncertainty in the flux measurement.}
\end{deluxetable}

The search observations were performed between 2010 November and 2013 September with the GMRT Software Back-end (GSB, \cite{roy10}) producing simultaneous incoherent and coherent beam filter-bank outputs of $512\times 0.0651$ MHz sampled every 61.44 $\mu$s. Details of the observational configuration are described in \cite{bh13}. Positional uncertainty associated with the \emph{Fermi}-LAT sources can easily be covered by the wider incoherent beam of the GMRT 
In addition to the wider incoherent beam, data from the more sensitive coherent beam 
were simultaneously recorded with much narrower beam width ($\sim$ $\pm$ 1.5\arcmin~at 322 MHz and $\sim$ $\pm$ 80\arcsec~at 607 MHz), which is useful in cases where the pulsar happens to be close enough to the pointing center. However, this is not likely considering the typical positional uncertainty ($\sim$ $\pm$10\arcmin) of the \emph{Fermi}-LAT sources.

Using parameters of 32 MHz bandwidth, 10\% duty cycle, incoherent array gain of 2.3 K/Jy, for a 30-minute observing time, we estimate the search sensitivity using radiometer equation \citep{lorimer04} for a 5$\sigma$ detection as (66K $+T_\mathrm{sky}$)/(335K) mJy at 322 MHz and (92K $+T_\mathrm{sky}$)/(335K) mJy at 607 MHz, where 66K and 92K are the receiver temperatures at the respective frequencies.
Thus considering $|b|>$ 5\degr~and $T_\mathrm{sky}$ $\sim$ 33--220 K at 322 MHz, minimum detectable flux at 322 MHz for 5$\sigma$ detection is 0.3--0.9 mJy. Whereas, considering $|b|>$ 5 \degr~and $T_\mathrm{sky}$ $\sim$ 10--45 K at 607 MHz, our search sensitivity for 5$\sigma$ detection at 607 MHz is 0.3--0.4 mJy. The sky temperature is estimated from the all-sky 408 MHz image by Haslam et al. (1982).  This sky temperature is then scaled to the observing frequency using an assumed spectral index of  $-$2.55 for the brightness temperature of Galactic synchrotron emission.

The data are processed in the NCRA HPC cluster and in the IUCAA HPC cluster with Fourier-based acceleration search methods using PRESTO \citep{ransom02}.  We investigated trial dispersion measures (DMs) ranging from 0 pc cm$^{-3}$ up to 350 pc cm$^{-3}$, which is the limiting DM for pulsars at $|b|>5^\circ$ up to distance of 8 kpc (according to NE2001, \cite{cl01}). Since we are observing at low frequencies we are only sensitive to nearby MSPs; at higher DMs the survey sensitivity decreases because of dispersive smearing within the channels. A linear drift
of up to 200 Fourier-frequency bins for the highest summed harmonic was allowed. The powerline, 50 Hz, and its subsequent
harmonics were excised. 
 
The newly discovered MSPs can be localised in the image plane with the GMRT interferometric array with an
accuracy of better than $\pm$ 10\arcsec (half of the typical synthesized beam used in the image made at 322 MHz) using gated imaging of pulsars (\cite{roy13}, see  Figure \ref{fig:J0248_img}) and the multipixel beam former \citep{roy12} which is detailed in Section \ref{sec:localisations}.
Once the MSPs are localised in the image plane, we use the coherent array for follow up observations with a smaller field of view but with enhanced sensitivity. Using the coherent array with the central core of the GMRT having 17 antennas
(i.e. gain of $\sim$ 7 K/Jy) the timing sensitivity is 0.3 mJy for a 10$\sigma$ detection. After discovery we started a
regular timing campaign at 322 and 607 MHz over $\sim$ 5 years.

\begin{table*}
\centering
{\scriptsize
\caption{Timing parameters of PSR J0248$+$4230, J1207$-$5050 and J1536$-$4948
\label{tab:residuals}}
\begin{tabular}{|l|c||c|c|c|}
\hline
Name &J0248$+$4230 &J1207$-$5050 &J1536$-$4948 \\
  \hline
  \multicolumn{4}{c}{Gated imaging position$^\ast$} \\
  \hline
 Right ascension (J2000)\dotfill & 02$^\mathrm{h}$48$^\mathrm{m}$29$^\mathrm{s}$(7) & 12$^\mathrm{h}$07$^\mathrm{m}$21$^\mathrm{s}$(5)  & 15$^\mathrm{h}$36$^\mathrm{m}$24$^\mathrm{s}$(10)\\
 Declination (J2000)\dotfill     &$+$42\degr30\arcmin13\arcsec(4)         & $-$50\degr50\arcmin30\arcsec(10) & $-$49\degr48\arcmin45\arcsec(10)\\
  \hline
  \multicolumn{4}{c}{Parameters from radio timing$^\ast$} \\
  \hline
 Right ascension (J2000)\dotfill &02$^\mathrm{h}$48$^\mathrm{m}$31\fs003(1)         &12$^\mathrm{h}$07$^\mathrm{m}$22\fs40392(7) & 15$^\mathrm{h}$36$^\mathrm{m}$23\fs22091(3)\\
 Declination (J2000)\dotfill     &$+$42\degr30\arcmin20\farcs49(5)                 & $-$50\degr50\arcmin38\farcs680(1) &$-$49\degr48\arcmin54\farcs6880(8) \\
 Proper motion in RA(mas yr$^{-1}$)\dotfill &$-$ & 6.9(4) &$-$ 7.3(2)\\
 Proper motion in DEC(mas yr$^{-1}$)\dotfill &$-$ &1.4(5) &$-$2.7(5)\\
 Frequency $f$ (Hz)\dotfill   & 384.49193525267(4)                           &206.493931730035(8) & 324.68438438109(6) \\
 Frequency derivative $\dot{f}$ (Hz s$^{-1}$)\dotfill &$-$2.4944(7)$\times$10$^{-15}$  &$-$2.586(2)$\times$10$^{-16}$ &$-$2.2338(1)$\times$10$^{-15}$\\
 Period epoch (MJD)\dotfill           &56588.0                                      &56478.0 & 56530.6\\
 Dispersion measure $\mbox{DM}$ (pc~cm$^{-3}$) &48.2634(1)        &50.67   & 38.00125(4)\\
 DM $1^{\mathrm st}$derivative $\mbox{DM1}$\dotfill & $-$    & $-$    & $-$0.00038(1)\\
 DM $2^{\mathrm nd}$ derivative $\mbox{DM2}$\dotfill & $-$    & $-$    & 0.000086(10)\\
 Binary model\dotfill & $-$                   &  & DDH\\
 Orbital period $P_{b}$ (days)\dotfill&$-$&$-$ & 62.05149821(2) \\
 Projected semi-major axis $x$ (lt-s) & $-$ & $-$ & 30.31454(4)\\
 Ascending node epoch $T_{\rm {ASC}}$ (MJD)\dotfill &$-$           &$-$ & 56580.194(7)\\
 Orthometric amplitude $h_3$ ($\mu$s) \dotfill &$-$           &$-$ & 9(4)\\
 Timing Data Span     \dotfill           &56375.3$-$58084.8 & 55728.6$-$57803.8 & 56062.7$-$58084.3\\
 Number of TOAs\dotfill  & 149              &158 & 788 \\
 Reduced Chi-square \dotfill       & 1.05        & 0.5 & 1.21\\
 Post-fit residual rms (ms)\dotfill       & 0.012                       & 0.010 & 0.010\\\hline
 \multicolumn{4}{c}{Derived parameters} \\
  \hline
 Period (ms) \dotfill                                         & 2.60083478563164(3)         & 4.84275732280274(2) & 3.07991405840531(6) \\
 Period Derivative (s/s) \dotfill                             & 1.6873(2)$\times$10$^{-20}$ & 6.0677(5)$\times$10$^{-21}$ & 2.1209(2)$\times$10$^{-20}$ \\
 Total time span (yr) \dotfill                                & 4.6                          & 5.6                &  11.5\\
 Energy loss rate $\dot{E}$ (erg/s)\dotfill                   & 3.8$\times$10$^{34}$        & 2.1$\times$10$^{33}$   & 2.9$\times$10$^{34}$\\
 $\dot{E}$ with kinematic corrections (erg/s)\dotfill & $-$ & 2.1$\times$10$^{33}$& 2.7$\times$10$^{34}$\\
 Characteristic age (yr)\dotfill                              & 2.4$\times$10$^{9}$          & 12.6$\times$10$^{9}$ & 2.3$\times$10$^{9}$\\
 Surface magnetic field (Gauss)\dotfill                & 2.1$\times$10$^{8}$          & 1.7$\times$10$^{8}$ & 2.5$\times$10$^{8}$\\
 DM distance (kpc)$^\ddagger$ \dotfill                        & 1.8                          & 1.5 & 1.8\\
 DM distance (kpc)$^\ddagger$$^\dagger$ \dotfill              & 2.5                          & 1.3 & 0.98\\
 Companion mass M$_{\odot}$      \dotfill                           & $-$                          & $-$ & 0.27$-$0.74 \\\hline
\end{tabular}
}\\
{\small $^\ast$ Errors correspond to 1$\sigma$.\\
$^\ddagger$ using the \cite{cl01} model of electron distribution\\
$^\ddagger$$^\dagger$ using the \cite{yao17} model of electron distribution\\
We note that the calculated DM distance is model dependent.\\
Timing uses DE421 solar system ephemeris.\\
The numbers in the parenthesis are uncertainties in preceding digits.}
\end{table*}

\begin{table*}
\centering
\caption{$\gamma-$ray results \label{tab:latres}}
\vspace{0.3cm}
\begin{tabular}{|l|c|c|c|}
\hline
Name   & PSR J0248+4230  & PSR J1207$-$5050 & PSR J1536$-$4948 \\
\hline
\multicolumn{4}{c}{Spectral fit results}\\
\hline
$N_{0}$ (10$^{-9}$ cm$^{-2}$ s$^{-1}$ GeV$^{-1}$)\dotfill & 0.4$\pm$0.2 & 1.2$\pm$0.4 & 7.2$\pm$0.2\\
$\Gamma$\dotfill & 0.6$\pm$0.6 & 0.9$\pm$0.3 & 1.6$\pm$0.1\\
$E_{C}$ (GeV)\dotfill & 1.5$\pm$0.6 & 1.7$\pm$0.4 & 6.7$\pm$0.4\\
$F$ (10$^{-9}$ cm$^{-2}$ s$^{-1}$)\dotfill & 1.1$\pm$0.5 & 4.5$\pm$1.1 & 80.1$\pm$1.4\\
$G$ (10$^{-12}$ erg cm$^{-2}$ s$^{-1}$)\dotfill & 1.6$\pm$0.3 & 5.6$\pm$0.6 & 80.7$\pm$1.4\\
TS\dotfill & 103 & 420 & 12152\\
$L_{\gamma \rm{NE2001}}$ (10$^{32}$ erg s$^{-1}$)\dotfill & 6.2$\pm$1.2 & 1.5$\pm$0.2 & 310$\pm$5 \\
$L_{\gamma \rm{YMW2017}}$ (10$^{32}$ erg s$^{-1}$)\dotfill & 12$\pm$2 & 11$\pm$1 & 93$\pm$2 \\
$\eta_{\gamma \rm{NE2001}}$ (\%)\dotfill & 1.6$\pm$0.3 & 70.5$\pm$7.7 & 114.8$\pm$1.9 \\
$\eta_{\gamma \rm{YMW2017}}$ (\%)\dotfill & 3.2$\pm$0.6 & 53.8$\pm$5.7 & 34.4$\pm$0.7 \\
\hline
\multicolumn{4}{c}{Pulse profile fitting results}\\
\hline
$\phi_{P1}$\dotfill & 0.151$\pm$0.015 & 0.272$\pm$0.006 & 0.196$\pm$0.002\\
$w_{P1}$\dotfill & 0.055$\pm$0.020 & 0.046$\pm$0.007 & 0.024$\pm$0.003\\
$\phi_{P2}$\dotfill & 0.830$\pm$0.019 & 0.775$\pm$0.004 & 0.453$\pm$0.001\\
$w_{P2}$\dotfill & 0.085$\pm$0.015 & 0.014$\pm$0.003 &  0.025$\pm$0.002\\
$\phi_{P3}$\dotfill & $-$ & $-$ & 0.647$\pm$0.001\\
$w_{P3}$\dotfill & $-$ & $-$ & 0.016$\pm$0.002\\
$\phi_{P4}$\dotfill & $-$ & $-$ & 0.788$\pm$0.001\\
$w_{P4}$\dotfill & $-$ & $-$ & 0.011$\pm$0.001\\
$\Delta$\dotfill & 0.679$\pm$0.024 & 0.503$\pm$0.007 & 0.591$\pm$0.002\\
$\phi_{B1}$\dotfill & $-$ &$-$ & 0.306$\pm$0.016\\
$w_{B1}$\dotfill & $-$ & $-$ & 0.110$\pm$0.015\\
$\phi_{B2}$\dotfill & $-$ & $-$ & 0.757$\pm$0.004\\
$w_{B2}$\dotfill & $-$ & $-$ & 0.063$\pm$0.004\\
\hline
\end{tabular}
{\small Note: The photon and energy fluxes reported in rows 4 and 5 of the spectral fit results are integrated from 0.1 to 300 GeV.  The point source test statistic (TS) values reported in row 6 of the spectral fit results are calculated as described in \citet{4FGL}.  The $\gamma$-ray luminosity ($L_{\gamma}$) and efficiency ($\eta_{\gamma}$) values \citep[see][]{2PC} are reported in rows 7 through 10 of the spectral fit results, with subscripts indicating which distance estimate was used.  For $\eta_{\gamma}$, we use the kinematically corrected $\dot{E}$ values, when available.  The uncertainties in rows 7 through 10 use the uncertainties on $G$ only. The $\Delta$ value listed in row 9 of the pulse profile fitting results is the difference in phase between the first and last $\gamma$-ray peak.  All uncertainties are statistical only.}
\vspace{1cm}
\end{table*}
\section{Discovery of three MSPs}
\label{sec:discovery}
PSR J0248+4230 was discovered in a 30$-$minute observing run with the GMRT at 322 MHz targeted at the $\gamma$-ray source 4FGL J0248.6+4230 \citep{4FGL}.
Here, and throughout the paper, we refer to the 4FGL names of the associated LAT sources even though our initial observations were targeted at sources from earlier source lists and catalogs.
It is a 2.60 ms MSP with DM of 48.25 pc cm$^{-3}$ having flux density of 7.5 mJy at 322 MHz and 0.96 mJy at 607 MHz. We estimate a 
spectral index of $-$3.34(6) for this pulsar. 
At 607 MHz, we could resolve two components 
of this MSP profile (Figure \ref{fig:multiprofiles}). The leading component has a significantly higher amplitude than the following component. Due to dispersion smearing, the profile components are not well resolved at 322 MHz. 

PSR J1207$-$5050 was discovered in a 30$-$minute observing run with the GMRT at 607 MHz, targeted at the $\gamma$-ray source 4FGL J1207.4$-$5050 \citep{4FGL}. It is a 4.84 ms pulsar with a DM of 50.67 pc cm$^{-3}$, having flux density of 0.5 mJy at 607 MHz. We did not detect this MSP at 322 MHz with the GMRT, even with observations at multiple epochs, which could be due to the fact that the profile is smeared at 322 MHz. Considering the sky temperature at the position of the pulsar, the 10$\sigma$ non-detection limit at 322 MHz is 0.38 mJy for 30$-$minute observing time using 17 antennas in phased array. Considering this limiting flux density we estimate spectral index of $\sim$0.43(6), indicating possible spectral turn over between 607 and 322 MHz, which is higher than seen for general MSP population (e.g. studies by \cite{kramer98,dai15}).
We aim to perform coherently dedispersed observations of this pulsar at 322 MHz to either detect or place more stringent limits on the emission.

PSR J1536$-$4948 was discovered in a 30$-$minute observing run with the GMRT at 322 MHz, targeted at the $\gamma$-ray source 4FGL J1536.4$-$4948 \citep{4FGL}. 
It is a 3.07 ms pulsar with a DM of 38.00 pc cm$^{-3}$ having flux density of 12 mJy at 322 MHz. We have also detected this MSP at 607 MHz having a flux density 4 mJy. We estimate a spectral index of $-$2.86(6) for this pulsar.
We observe a very wide pulse profile (profile width $>$ 350\degr) with 3 components, but due to dispersion smearing ($\sim$ 20\% of pulse period) the profile components 
are not well resolved at 322 MHz. Table \ref{discovery} summarizes the discovery parameters of these three MSPs.   

\section{Localization of the MSPs}
\label{sec:localisations}
Following the discoveries with the GMRT incoherent array 
(half power beam width $\sim$ 80\arcmin~for 322 MHz, 40\arcmin~for 607 MHz), 
using the 
techniques of multi-pixel beam-forming \citep{roy12} and MSP gating correlator \citep{roy13}, we could significantly 
improve the large positional uncertainties allowing us to use the sensitive coherent array (4 to 5 times more than incoherent array for the GMRT) for the follow-up timing observations which substantially reduces the use of array telescope time (16 to 25 times for the GMRT).

Since PSR J0248$+$4230 is a relatively weak MSP, we were not expecting to get significant signal-to-noise in the continuum image plane. Because the radio pulse profiles MSPs 
are dispersion broadened, specially for this MSP with wide emission components, we used a coherently dedispersed gating correlator, with proper optimisation, when selecting average on and off visibility phase bins. This MSP is unambiguously detected in a 30$-$minute observing run with 13$\sigma$ detection significance in an on$-$off gated image (Figure \ref{fig:J0248_img}). The pulsar is found as the only point source in the image with the precise position being 2$^\mathrm{h}$48$^\mathrm{m}$29$^\mathrm{s}$(7), $+$42\degr30\arcmin13\arcsec(4). 

\cite{roy13} reported a precise position for PSR J1207$-$5050 using a coherently dedispersed gated correlator. Since the radio pulse profile of PSR J1536$-$4948 is wide, we used a multi-pixel beamformer to determine its precise position \citep{roy12}. 
Upper part of the Table \ref{tab:residuals} lists the precise astrometric positions of these three MSPs. These precise positions were then used for follow-up timing observations with the coherent array mode of the GMRT detailed in Section \ref{sec:timing}.

\section{Timing study}
\label{sec:timing}
Following the precise astrometric localization of these three MSPs described in Section \ref{sec:localisations}, we conducted dense observations using the GMRT coherent array to time the MSPs.  This timing campaign allowed us to construct timing models that describe well the pulse times of arrival (TOAs). These timing models can be extended to nearby epochs, and allow us to start observing the MSPs more sparsely. We used the highest signal$-$to$-$noise ratio profiles as templates for extracting TOAs. For PSR J0248$+$4230 and J1536$-$4948, the follow-up observations were at 322 and at 607 MHz, whereas for PSR J1207$-$5050 the follow up observations were at 607 MHz as this pulsar was not detected in 322 MHz observations, as discussed in Section \ref{sec:discovery}.

Both PSR J0248$+$4230 and PSR J1207$-$5050 are isolated pulsars, making it easier to derive phase-connected solutions. For PSR J1536$-$4948, obtaining a phase-coherent timing solution was more difficult. During the phase connection procedure we reached a point where the number of rotations between any remaining observations was ambiguous. Although this situation normally happens for objects with sparse observations, our observations of PSR~J1536$-$4948 have a relatively good cadence. The issue is caused instead by the relative faintness of the pulsar and the broad pulse profile (see Figure~\ref{fig:multiprofiles}), which results in a low timing precision. Furthermore, too many parameters (spin, astrometric and orbital, all on similar timescales) can be adjusted when fitting for timing delays.
Because of this, we had to use the algorithm described by \cite{fr18}\footnote{\url{https://github.com/pfreire163/Dracula}}, which
uses the {\sc tempo}\footnote{\url{http://tempo.sourceforge.net/}} timing software to
explore all the combinations of rotation numbers between unconnected observations that result in timing solutions with a low residual $\chi^2$. The algorithm soon determined the correct set of rotation numbers between all observations, which yields the timing solution.

The timing solutions from $\sim$ 5 years of observations with the GMRT, derived using standard 
pulsar timing software {\sc tempo2}\footnote { \url {http://www.atnf.csiro.au/research/pulsar/tempo2} }, are presented in
Table \ref{tab:residuals}. The timing residuals (the observed TOAs minus the prediction of the model for the TOAs)
are displayed in Fig.~\ref{fig:residual_J0248}, \ref{fig:residual_J1207} and \ref{fig:residual_J1536},
for PSR J0248$+$4230, J1207$-$5050 and ~J1536$-$4948 respectively. The residual
root mean squares (rms) are 10, 12 and 10$\,\mu$s respectively. 

For timing of PSR~J1536$-$4948, we use not only the radio data, but also TOAs derived from LAT $\gamma$-rays (see section~\ref{subsec:gammapulses}). The residuals showed a small secular drift between the radio and $\gamma$-ray TOAs, caused by long-term change in DM, which is due to the relative motion of the pulsar and the Earth changing the column density of ionized gas between both. This variation of the DM is confirmed by comparison of the radio TOAs at frequencies of 322 and 607 MHz, and can be modeled well with two DM derivatives, which are also listed in Table~\ref{tab:residuals}.

We describe the orbital motion of PSR J1536$-$4948 using the DDH model \citep{fw10}, which is based on the earlier DD model \citep{dd86} but re-parameterised to yield less-corelated Shapiro delay parameters. 
The mass function of PSR J1536$-$4948 is 0.007768 M$_\sun$, which, assuming a pulsar mass of 1.4 M$_\sun$ and orbital inclinations of 90, 60 and 25\degr~ would yield companion masses of 0.27, 0.32 and 0.74 M$_\sun$ respectively. We have not seen any evidence of eclipsing from this wide binary, suggesting that the companion is a Helium white dwarf (WD). For the orbital period of this system, the expectation of the \cite{ts99} model is a Helium WD mass of $\sim\, 0.3\, M_\sun$; which suggests that the orbital inclination is close to the median of $60^\circ$.

The DDH fit provides a weak (2$\sigma$) detection of the orthometric amplitude of the Shapiro delay ($h_3$).
In the absence of other post-Keplerian parameters, this is not enough for a determination
of the mass of the pulsar or the mass of the companion, nor of the orbital inclination. 
We also note that the current timing precision (with the observations using the GMRT Software Back-end having 32 MHz bandwidth) is not sufficient to make reliable estimation of the Shapiro delay. Ongoing observations with the upgraded GMRT wide band system \citep{gupta17} will allow to increase the timing span with more precise TOAs for better estimation of Shapiro delay.

Pulsar distance estimates come from comparing observed DM with the Galactic electron density $n_e(\vec{x})$ integrated along the line of sight (LoS) to the pulsar, using the models for  $n_e(\vec{x})$ provided by \cite{yao17} and \cite{cl01}. Comparing the DM distances with those obtained by other methods suggests that for most pulsars, the uncertainty is roughly gaussian, with a standard deviation near 30\%. Unfortunately, the difference distribution has very broad tails: for some pulsars the disagreement is a factor of a few to several. The tools described in \cite{Theureau11} and \cite{Hou14} attempt to identify the aberrant cases by comparing the models with HI, CO, and H$_\alpha$ observations.

We examined the LoS to our three pulsars to see if they might cross unmodeled electron over- or under-densities.  Nothing is unusual for PSRs J0248$+$4230 and J1536$-$4948, meaning that their DM distances are probably reliable. PSR~J1207$-$5050, however, has  noteworthy features. The first is that the LoS crosses the edge of the  Local Bubble high electron density region. The model geometry is highly idealized, and the predicted $\sim 20$ pc cm$^{-3}$ step could easily be off by a factor of two, which could shift the distance by $\pm$ a few hundred pc. Next is that \cite{yao17} highlight PSR~J1227$-$4853 as being a pulsar for which the model distance is  ~600 pc less than that obtained from optical photometry of its binary companion (\cite{dcm+14}). 
PSR~J1207$-$5050 is in the same part of the sky, with about the same DM, and nominally at about the same distance, and PSR~J1227$-$4853 has a  similar $\sim 10$ pc cm$^{-3}$ step due to the edge of the Local Bubble. But the 600 pc discrepancy appears to not exist: \cite{jkc+18} show that the Gaia parallax measurements of PSR J1227--4853's optical companion give a distance  matching both DM distances. This would bolster confidence in PSR~J1207$-$5050's DM distance, except that its LoS passes near two B2 stars, Hipparcos 59173 and 59196, both within 130 pc of Earth. These hot stars can completely ionize the interstellar medium within a few tens of parsecs, depending on the unknown local gas density, creating unmodeled electrons. Indeed, the $H_\alpha$ maps of \cite{Finkbeiner03} show that PSR~J1207$-$5050 lies at the edge of a bright $H_\alpha$ glow, presumably created by the stars and thus well in the pulsar's foreground.  Unmodeled extra electrons would mean that the pulsar is closer than predicted by the electron models. This does not seem to be the case, because the ratio of both observed $H_\alpha$ intensity and calculated emission measure at the two pulsar positions is the same. We conclude that the DM distance to PSR~J1207$-$5050 with a 30\% uncertainty is likely to be correct.

\section{Gamma-ray detection}
\label{sec:gammaray}
To confirm identification of the newly detected radio MSPs with the corresponding unassociated LAT sources, we need to detect significant pulsations at the spin period in the $\gamma$-ray data.  We therefore performed spectral and timing analyses of the LAT data, using the radio timing solutions, as described below.

\subsection{LAT data preparation}
\label{subsec:latdataprep}
We analyzed LAT Pass 8 data \citep{atwood13,bruel18} within 15$^{\circ}$ of the best-fit radio timing position of each MSP, separately, starting from the beginning of the mission, 2008 August 4, and ending 2020 April 27.  We kept all events with reconstructed energies from 0.05 to 500 GeV, zenith angles less than 90$^{\circ}$, and belonging to the \texttt{SOURCE} event class.  We filtered the data to create good time intervals when the spacecraft was in nominal science operations mode, the data were flagged as good, and to avoid LAT-detected solar flares and gamma-ray bursts.

\subsection{Gamma-ray spectral analysis and results}
\label{subsec:specanalysis}
We created spatial and spectral models of the regions around each MSP using the \emph{Fermi}-LAT Fourth source catalog \citep[4FGL,][]{4FGL}, including all sources within 25$^{\circ}$ of the pulsar and the corresponding diffuse emission components. For all three MSPs, the position of the associated 4FGL source was $\leq1.2\arcmin$ from the timing position, consistent within the 4FGL positional uncertainty.  We chose to move the 4FGL source associated with each MSP (4FGL J0248.6+4230, 4FGL J1207.4$-$5050, and 4FGL J1536.5$-$4948) to the radio timing position.  The $\gamma$-ray spectrum of each source associated with one of our MSPs was modeled using an exponentially cutoff power-law shape of the form described in Eq.~\ref{eq:plec}, observed to describe the spectra of most $\gamma$-ray pulsars well \citep{2PC}.  
\begin{equation}\label{eq:plec}
\frac{dN}{dE}\ =\ N_{0}\Big(\frac{E}{E_{0}}\Big)^{-\Gamma}\exp \Big\lbrace-\Big(\frac{E}{E_{C}}\Big)^{b}\Big\rbrace
\end{equation}

In Eq.~\ref{eq:plec}, $N_{0}$ is a normalization parameter with units of GeV$^{-1}$ cm$^{-2}$ s$^{-1}$ and is calculated from the 4FGL information to be the value of the differential counts spectrum at the pivot energy $E_{0}$, $\Gamma$ is the low-energy photon index, $E_{C}$ is the cutoff energy, and $b$ is an exponential index controlling how quickly the spectrum cuts off.  We chose to fix $b$ to a value of 1, but did explore other values, as discussed later in this section.

For each MSP, we performed a binned maximum likelihood fit, using the \texttt{P8R3\_SOURCE\_V2} instrument response functions\footnote{See \url{https://www.slac.stanford.edu/exp/glast/groups/canda/lat_Performance.htm}.}, in which we allowed the spectral parameters to vary for all sources within 6$^{\circ}$ of the pulsar that were found to have an average significance of $\geq10\sigma$ in the 4FGL catalog.  The spectral parameters of the Galactic and isotropic diffuse emission components were also allowed to vary in the fits.  For sources not meeting the previous criteria which were flagged as significantly variable in the 4FGL catalog, we allowed their spectral normalizations to be free in the fits if they were within 8$^{\circ}$ of the corresponding pulsar position.  The spectral analysis was done over the energy range of 0.1 to 300 GeV but the exposure products were calculated over the entire energy range of our data, with 10 bins per decade, to allow for the use of energy dispersion\footnote{See \url{https://fermi.gsfc.nasa.gov/ssc/data/analysis/documentation/Pass8_edisp_usage.html}.}.

After an initial fit, we examined the spatial residuals to determine if the spectral parameters of any sources we had fixed to the 4FGL values needed to be allowed to vary and if there was evidence for new sources not in the 4FGL catalog.  In doing so, for the region around PSR J1536$-$4948 we decided that the spectral normalization of the point source 4FGL J1457.3$-$4246, $9\fdg7$ from the MSP position, needed to be set free and redid the fit.

The 4FGL catalog uses a different spectral parameterization of the exponentially cutoff power-law shape\footnote{See the entry for the \texttt{PLSuperExpCutoff2} model at \url{https://fermi.gsfc.nasa.gov/ssc/data/analysis/scitools/source_models.html}.} for known $\gamma$-ray pulsars and fixes the $b$ parameter to a value of 2/3 based on what is observed in the spectra of the brightest $\gamma$-ray pulsars.  In our fitting, we found that the $\Gamma$ parameter for PSR J0248+4230 was unstable and often fit to $\approx$ 0.  When we instead set $b$ to a value of 1, the $\Gamma$ parameter was more well-behaved, but we found a strong dependence on the starting value of other parameters and chose to instead switch to the formulation given in Eq.~\ref{eq:plec}.  To be consistent, we used this spectral shape for all three MSPs.

We performed likelihood analysis with the $b$ parameter free, as well as fixed to the value of 2/3 used in the 4FGL catalog.  For PSRs J0248+4230 and J1207$-$5050, there was no significant change in the likelihood for $b=2/3$ or $b$ free, when compared to fits with $b=1$.  For PSR J1536$-$4948, the likelihood marginally favored a lower value of $b$, but the final result was very dependent on the starting value of $E_{C}$, possibly due to issues with modeling the diffuse emission in this region, so we chose to use and report only the $b=1$ results.

The resulting best-fit spectral parameters for each MSP are given in Table \ref{tab:latres} as well as the derived integral photon and energy flux values, $F$ and $G$, respectively.  Our best-fit parameters are not directly comparable to those reported in the 4FGL catalog, due to the differences in functional form, for the sources associated with these MSPs.  However, our fits with the same spectral model were and the results in Table \ref{tab:latres} yield compatible values of $F$ and $G$.

\subsection{Gamma-ray pulsation detection and timing}
\label{subsec:gammapulses}
Using the best-fit models of the regions, we selected events within 3$^{\circ}$ of each MSP with energies from 0.1 to 300 GeV and calculated spectral weights representing the probability that each event came from the pulsar of interest.  Use of these weights with the H test \citep{deJager89,deJager10} has been shown to enhance the sensitivity of searches for $\gamma$-ray pulsations in LAT data \citep{Kerr11}.  The resulting weighted H-test values  resulted in significant detections of $\gamma$-ray pulsations from all three pulsars, with values of 60.4 ($6.6\sigma$), 188.8 ($12.3\sigma$), and 3780.4 (60.4$\sigma$) for PSRs J0248+4230, J1207$-$5050, and J1536$-$4948, respectively.

For PSR J1536$-$4948, inspection of the pulse phase over time suggested that the timing solution did not extrapolate well before the radio discovery.  Following \citet{Ray11}, we used the 3$^{\circ}$ radius selection to construct $\gamma$-ray TOAs which were combined with the radio TOAs (as discussed in Section \ref{sec:timing}) to produce an improved timing model, yielding a weighted H-test result of 4644.5 (67.1$\sigma$), which indicates a significant improvement; this removed the drift in phase seen in the early LAT data.  The resulting $\gamma$-ray pulse profiles, as well as the radio profiles, are shown in Figure \ref{fig:multiprofiles}.

\subsection{Gamma-ray pulse profile characterization}
\label{subsec:gammaprof}
Once we had the final timing solutions\footnote{These timing solutions will be made available at \url{https://fermi.gsfc.nasa.gov/ssc/data/access/lat/ephems/}.}, we fit the $\gamma$-ray pulse profiles using the maximum likelihood method described in \citet{2PC} but restricted our model to Gaussian functional forms for the peaks.  The resulting fits are shown on the right hand side of each panel in Figure \ref{fig:multiprofiles} as the solid red line.  The fit values are given in Table \ref{tab:latres} where we have labeled the peaks in the order in which they appear in phase.
 
For PSRs J0248+4230 and J1207$-$5050, we found two Gaussian peaks described the data well.  The fit for J1207$-$5050 includes a substantial unpulsed component, visible above the estimated background level in the middle panel of Figure \ref{fig:multiprofiles}.  We tested adding additional, broad components to model this emission, but found it is best described by a simple, unpulsed pedestal comprising $\sim45$\% of the total emission.  Although these results depend on accurate photon weights, the quality of the spectral fit is good, and the excess emission is present even at higher energies where the LAT is more capable of distinguishing sources from backgrounds.  This component thus likely represents bona fide, nearly constant emission from the magnetosphere.  This measurement comes directly from the maximum likelihood fit to the photon weights, but is in good agreement with the background level drawn in Figure \ref{fig:multiprofiles}, which is an empirical estimator following the prescription in Section 5.1 of \citet{2PC}.
 
Fitting the $\gamma$-ray pulse profile for PSR J1536$-$4948 required six Gaussians, one for each of the four obvious peaks, and an additional Gaussian to account for the bridge emission between the first two peaks and the last two peaks.  The profile fit of this MSP also includes an unpulsed component accounting for $\sim7$\% of the pulsar emission, but given the numerous peaks spanning most of the pulse phase, it is possible this might just reflect wings/tails of the peaks which are not entirely fit with Gaussians.

\begin{figure}
\begin{center}
\includegraphics[width=6in,angle=0]{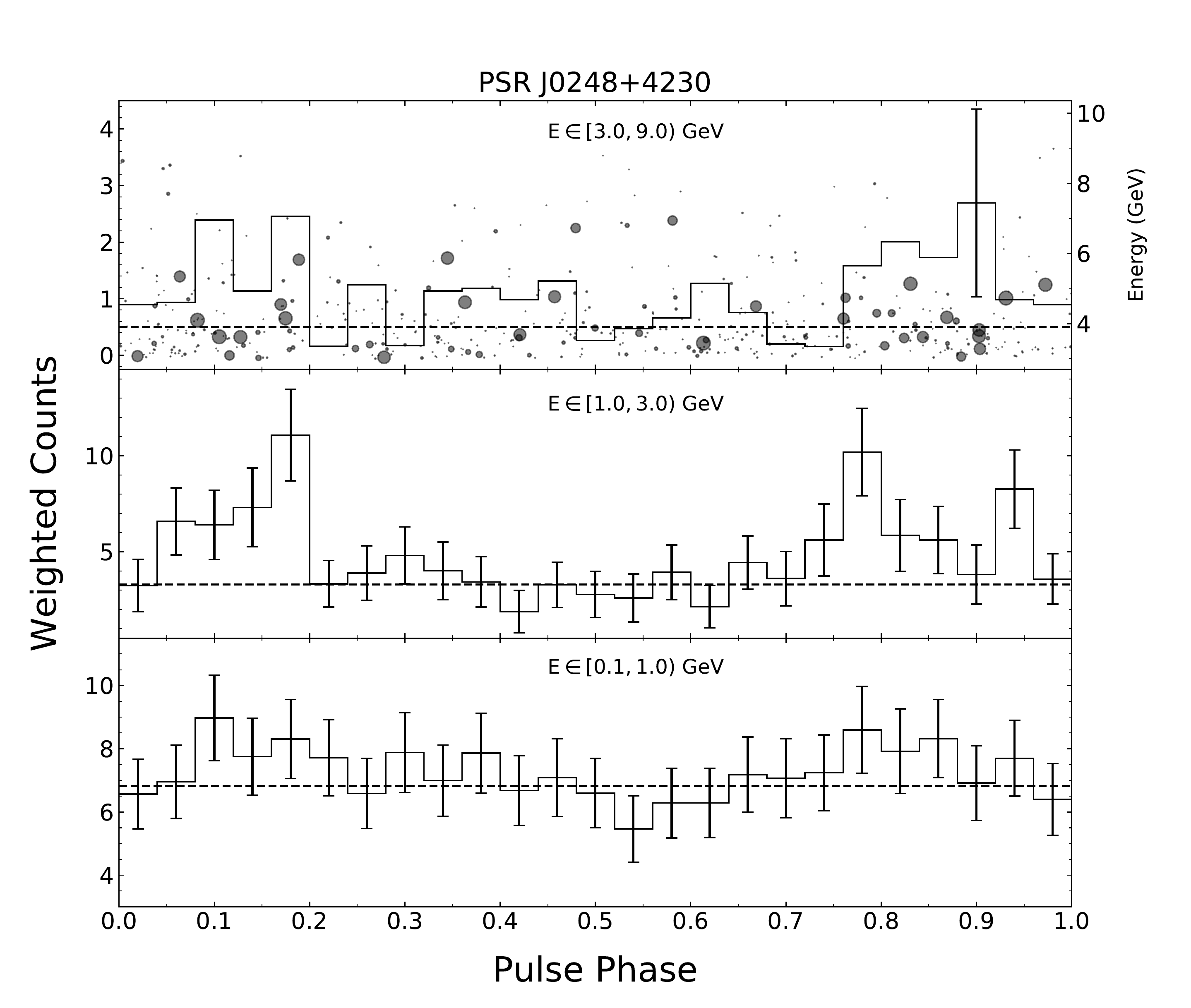}
\caption{The $\gamma$-ray pulse profile of PSR J0248+4230 in multiple energy bands, as indicated in the plot.  The dashed horizontal lines in each panel indicate the estimated background \citep[derived as in][]{2PC}.  In the top panel, we show only one, representative error bar, the rest are smaller than this by as much as 40\%.  The scatter plot in the top panel shows the phases and energies, right y-axis, of individual events with marker sizes proportional to the spectral weight values.}
\label{fig:ebandsJ0248}
\end{center}
\end{figure}

\begin{figure}
\begin{center}
\includegraphics[width=6in,angle=0]{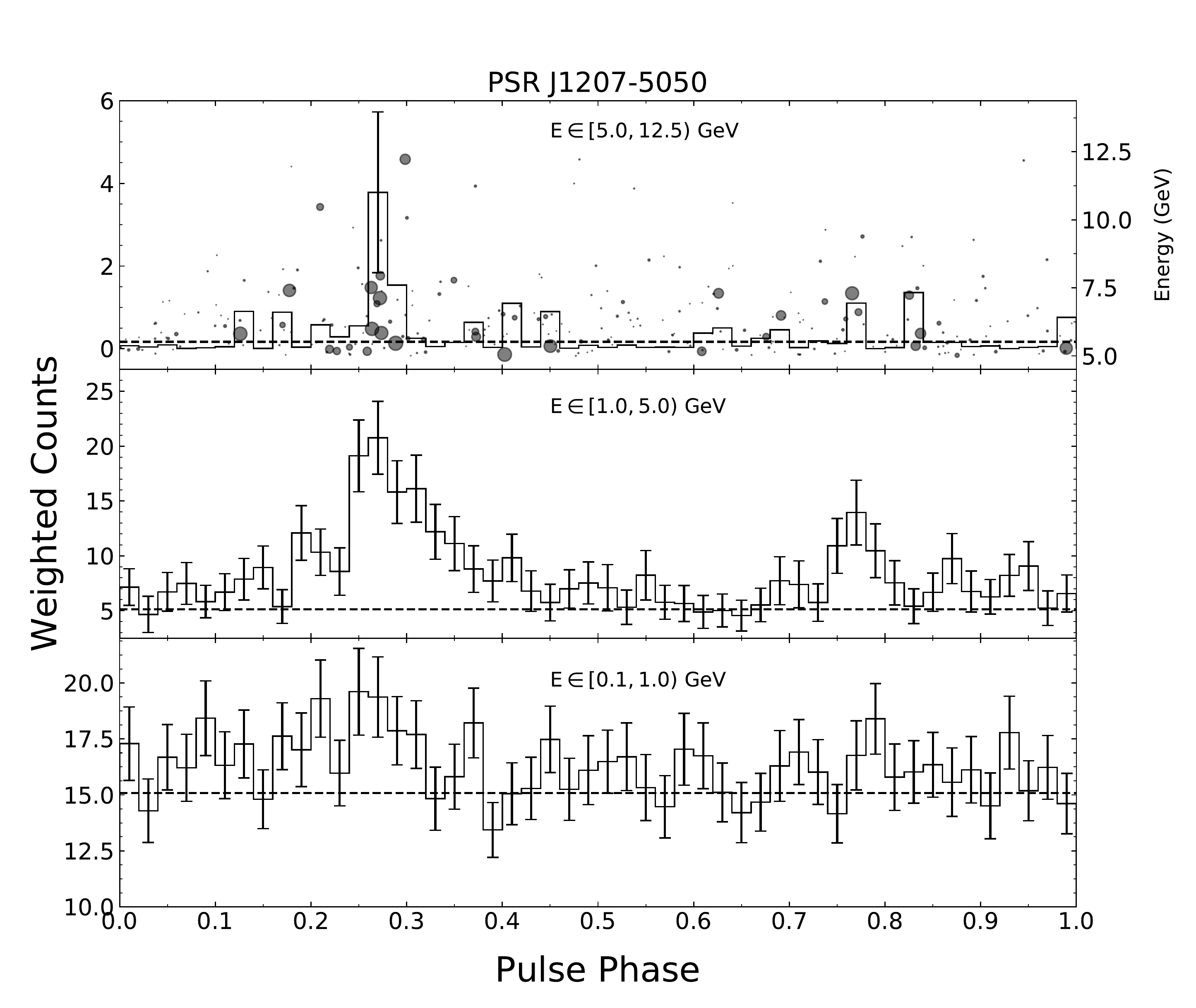}
\caption{The $\gamma$-ray pulse profile of PSR J1207$-$5050 in multiple energy bands, as indicated in the plot.  The dashed horizontal lines in each panel indicate the estimated background \citep[derived as in][]{2PC}.  In the top panel, we show only one, representative error bar, the rest are smaller than this by as much as 50\%.  The scatter plot in the top panel, the scatter plot shows the phases and energies, right y-axis, of individual events with marker sizes proportional to the spectral weight values..}
\label{fig:ebandsJ1207}
\end{center}
\end{figure}

\begin{figure}
\begin{center}
\includegraphics[width=6in,angle=0]{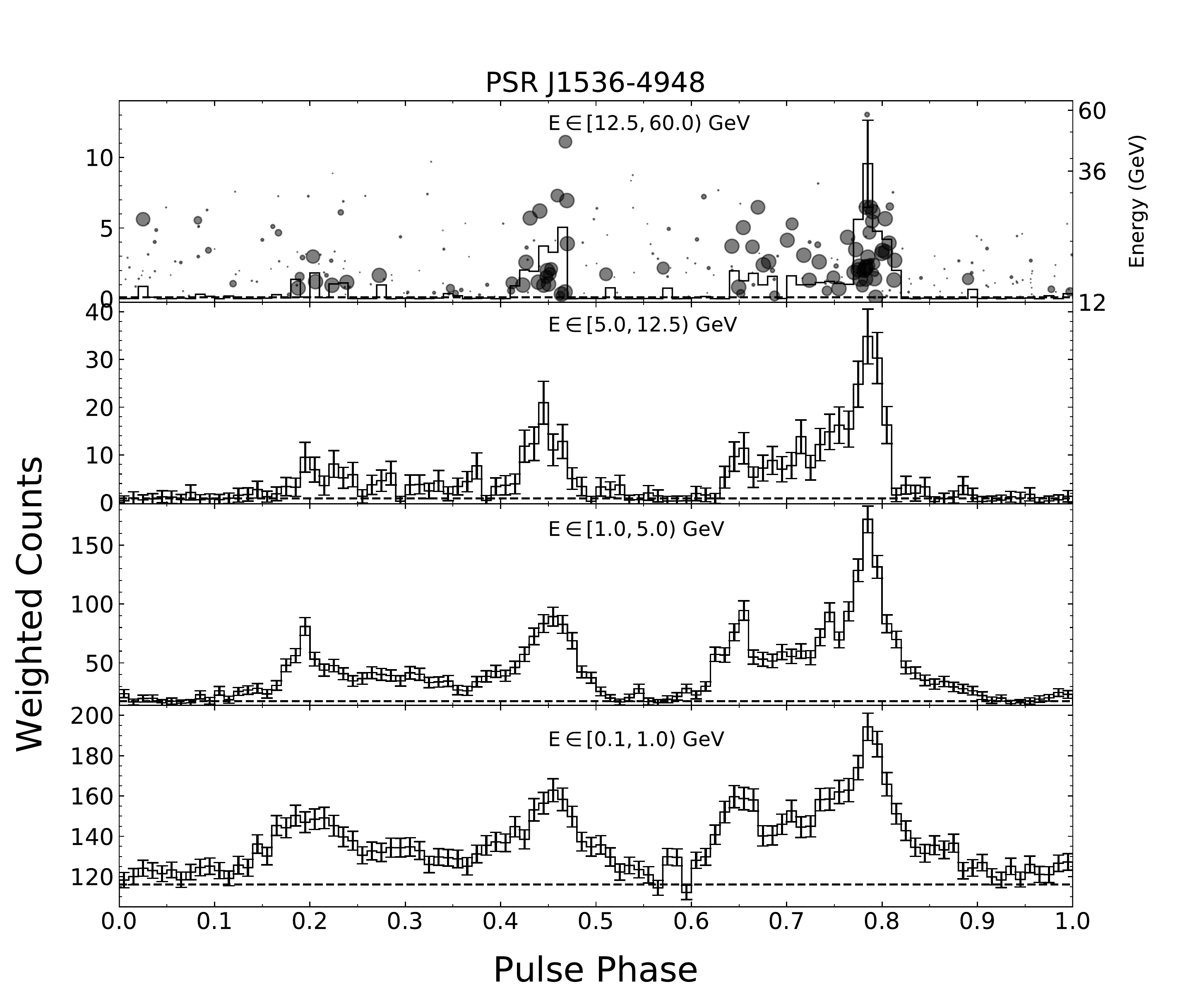}
\caption{The $\gamma$-ray pulse profile of PSR J1536$-$4948 in multiple energy bands, as indicated in the plot.  The dashed horizontal lines in each panel indicate the estimated background \citep[derived as in][]{2PC}.  In the top panel, we show only one, representative error bar, the rest are smaller than this by as much as 70\%.  The scatter plot in the top panel, the scatter plot shows the phases and energies, right y-axis, of individual events with marker sizes proportional to the spectral weight values.  Note that the energy axis in the top panel has a logarithmic scale.}
\label{fig:ebandsJ1536}
\end{center}
\end{figure}

\subsection{Pulse Profile Energy Evolution}
\label{subsec:heprofs}
Similar to what is observed at radio wavelengths, $\gamma$-ray pulse profiles can show interesting evolution when the data are split into smaller energy bands.  Figures \ref{fig:ebandsJ0248}, \ref{fig:ebandsJ1207}, and \ref{fig:ebandsJ1536} show the $\gamma$-ray pulse profiles in different energy bands for PSRs J0248+4230, J1207$-$5050, and J1536$-$4948, respectively.  For each MSP, to estimate the highest pulsed photon energy, we searched through the events within 3$^{\circ}$ of the radio position, requiring the spectral weight to be $\geq$ 0.001 and that the reconstructed event direction be compatible with the MSP position within the 95\% containment radius of the point-spread function for that energy, conversion layer, and incident angles with respect to the LAT instrument coordinates.

For both PSR J0248+4230 and J1207$-$5050, the bulk of the signal appears to come from the middle energy band, just above 1 GeV.  This is not surprising given the low values of $\Gamma$ (indicating a hard, power-law spectrum below the cutoff energy) and values of $E_{C}$ near 1 GeV in Table \ref{tab:latres}.  PSR J1536$-$4948, on the other hand, appears to show pulsations out to several tens of GeV.

With significant detection of $\gamma$-ray pulsations from the Vela and Crab pulsars above 100 GeV and even out to TeV energies \citep[e.g.,][]{CrabVERITAS,CrabMAGIC,VelaHESSII}, there has been growing interest in determining which other $\gamma$-ray pulsars might have detectable pulsations at energies out to 100 GeV or beyond \citep[e.g.,][]{SazParkHEpsrs}.  In order to assess the possibility for detection of PSRs J0248+4230, J1207$-$5050, and J1536$-$4948 at higher energies, we have included a scatter plot of individual event energies in the highest energy bands (top panels) of Figures \ref{fig:ebandsJ0248}, \ref{fig:ebandsJ1207}, and \ref{fig:ebandsJ1536}.  For these scatter plots, the size of the marker is proportional to the spectral weight.

For PSR J0248+4230, we found an event compatible with the MSP at phase 0.004 with an energy of 8.6 GeV and spectral weight (reflecting the probability of having come from that source) of 0.004.  This event is not likely to be associated with the pulsar, based on the low weight and the recorded phase not lining up with any observed feature in the pulse profile.  If we constrain our search to the pulse profile peaks (phases $\phi\in[0.0,0.2]\cup[0.7,1.0)$), we find an event at phase 0.189 with an energy of 5.8 GeV and a spectral weight of 0.633.  Events in the highest energy band have spectral weights ranging from 0.006 to 0.918.

For PSR J1207$-$5050, we found an event compatible with the MSP at phase 0.298 with an energy of 12.2 GeV and a spectral weight of 0.506.  This event falls within the phase range of the first peak and, based on the value of the weight, could plausibly be from the pulsar.  Events in the highest energy band have spectral weights ranging from 0.008 to 0.948.

For PSR J1536$-$4948, we found an event compatible with the MSP at phase 0.785 with an energy of 57.9 GeV and a spectral weight of 0.107.  While this event does occur at a phase compatible with the tallest peak in the $\gamma$-ray pulse profile, it does have a low weight.  This could be due, in part, to the choice of spectral model.  The next highest energy event occurs at phase 0.468, within the second tallest peak, and has an energy of 46.1 GeV and a spectral weight of 0.757.  Events in the highest energy band have spectral weights ranging from 0.107 to 0.996.  Taking events corresponding to the top panel of Figure \ref{fig:ebandsJ1536}, we calculate a weighted H test value of 220, corresponding to a detection of 13$\sigma$.

To further quantify evidence for pulsed emission above 10 GeV, we performed an analysis similar to that reported in \citet{1FHL} and \citet{SazParkHEpsrs}.  For each pulsar, we selected all events with energies from 1 to 10 GeV and with spectral weights $\geq$0.1 We used these events to generate a pulse profile that serves as a lower energy ``template''. We then selected all events with energies $\geq$10 GeV and reconstructed directions consistent with the respective pulsar within the 95\% containment radius of the point-spread function (0$\fdg$5 for events converting in the first of the LAT and 0$\fdg$8 for events converting in the back) and performed a likelihood test to determine if they were likely to come from a similar distribution as the 1 to 10 GeV template.  Following \citet{1FHL}, we required a tail probability, or p-value, of 0.05 to claim evidence for emission at higher energies.

For PSRs J0248+4232 and J1207$-$5050, we did not find significant evidence for pulsed emission above 10 GeV, in agreement with conclusions drawn from Figs. \ref{fig:ebandsJ0248} and \ref{fig:ebandsJ1207}.  For PSR J1536$-$4948, the likelihood results returned a p-value of 1.5$\times10^{-13}$ for events above 10 GeV, suggesting significant pulsed emission above this energy.  When applying the same analysis to events above 25 GeV for this pulsar, the likelihood test yields a p-value of 0.02, suggesting there is significant pulsed emission even above 25 GeV.  These results are in agreement with those of \citet{SazParkHEpsrs}, who claimed evidence for pulsed $\gamma$-ray emission above 25 GeV for five MSPs, including PSR J1536$-$4948, using a preliminary timing solution.

For the  third catalog of hard \emph{Fermi}-LAT sources (3FHL), \citet{3FHL} analysed 7 years of LAT data above 10 GeV looking for hard sources and associated this pulsar with the source 3FHL J1536.3$-$4949, based on positional coincidence only as a long-term timing solution had not yet been constructed.  Fitting data from 10 to 300 GeV, modeling the spectrum PSR J1536$-$4948 as a simple power law, we find an integrated photon flux of (2.4$\pm$0.2)$\times$10$^{-10}$ cm$^{-2}$ s$^{-1}$, energy flux of (6.5$\pm$0.8)$\times$10$^{-12}$ erg cm$^{-2}$ s$^{-1}$, and photon index of 3.5$\pm$0.3.  The photon index is the same as that found in the 3FHL catalog but our flux values are both lower, though the energy flux values agree within statistical uncertainties.  This may be due to the different versions of the Pass 8 data and diffuse models used between the two analyses. Therefore, we can now identify 3FHL J1536.3$-$4948 as PSR J1536$-$4948.

\section{Discussion}
\label{sec:Discussion}
An important, derived quantity for understanding high-energy pulsar emission is the $\gamma$-ray luminosity $L_{\gamma} = 4\pi d^{2} f_{\Omega} G$, where $f_{\Omega}$ is a beaming factor typically assumed to be near 1 \citep{2PC}.  From this luminosity, we can calculate the efficiency with which rotational energy is turned into $\gamma$-rays as $\eta_{\gamma} = L_{\gamma}/\dot{E}$.  Table \ref{tab:latres} reports values for $L_{\gamma}$ and $\eta_{\gamma}$ using both the distance estimate from \citet{cl01} and \citet{yao17}, and the values of $\dot{E}$ with kinematic corrections, when available.  Known $\gamma$-ray MSPs span a large range in $\eta_{\gamma}$ \citep{2PC}, from 1\% to $>$100\%, and these three MSPS are no different.  In the Second Catalog of LAT $\gamma$-ray Pulsars \citep{2PC}, only three MSPs have a higher $L_{\gamma}$, using the NE2001 distance, than PSR J1536$-$4948 (PSRs J0218+4232, J0614$-$3329, and J1823$-$3021A).  Of these three, two have values of $\eta_{\gamma}$ $<$20\%, but J0614$-$3329, like J1536$-$4948, has $\eta_{\gamma} >$100\%.  Efficiencies above 100\% might suggest values of $f_{\Omega}$ less than 1, that the distances used are overestimated, or that the ``standard'' values of neutron star mass and radius used may not be representative (e.g., using a radius of 14 km instead of 10 km will increase $\dot{E}$ by a factor of 2).

The prevailing models of $\gamma$-ray emission from rotation-powered pulsars posit that particles are accelerated along magnetic field lines near, or beyond, the light cylinder radius ($c$P/$2\pi$, where co-rotation with the star requires moving at the speed of light $c$).  Particle acceleration may happen within the light cylinder in relatively narrow vacuum gaps above the last open field line \citep[e.g.,][]{Cheng86,MH04} or over the full open volume above the polar cap \citep[e.g.,][]{Harding05}. Alternatively, the emission may originate outside the light cylinder in a striped wind \citep[e.g.,][]{SWind} or in regions near an equatorial current sheet \citep[e.g.,][]{KHK14}.  Features in the predicted pulse profiles depend on the assumed structure of the magnetosphere used in each model. Thus, testing the models is one way to better understand the complex magnetic fields of neutron stars.

Fitting the observed $\gamma$-ray pulse profiles using one of these models is one method of estimating the viewing geometry of the system, namely the inclination angle of the magnetic axis and the observer viewing angle, both relative to the spin axis \citep[e.g.,][]{Johnson14,CZ19}.  When combined with profiles at other wavelengths or geometry constraints from different methods, these fitting methods can be useful tests of the different emission models. The pulse profiles of PSRs J0248+4232, J1207$-$5050, and J1536$-$4948 all have interesting features which will serve as useful tests for emission models.

For most $\gamma$-ray pulsars, the main radio peak is recorded at an earlier phase than the first $\gamma$-ray peak.  A small fraction of MSP $\gamma$-ray pulse profiles, however, are observed to have their first peak occur before the radio.  This is predicted by models in which the full open volume above the polar cap is available for particle acceleration \citep{Harding05}.  The $\gamma$-ray profile for PSR J0248+4232 might fall into this category if we consider what we have called the second peak in Table \ref{tab:latres}, based on the order they appear with our choice of phasing, to be the first peak.  However, the models that predict that the $\gamma$-ray peak should precede the radio also predict much broader peaks and have difficulty matching the relatively sharp peaks we observe \citep{Venter09,Johnson14}.

The $\gamma$-ray pulse profile of PSR J1207$-$5050 looks typical at first glance \citep[compared to the many profiles in][]{2PC}.  The behavior of the two peaks with increasing energy, however, is the opposite of what is often seen.  In particular, usually the second, taller peak persists out to higher energies but with this source we observe that happening with the first peak in Figure \ref{fig:ebandsJ1207}.  These widely separated peaks are easily produced in most models \citep[see, for instance,][]{Venter09,Johnson14}, but require a large impact parameter, the difference between the magnetic inclination and viewing angles, which is difficult to reconcile with the detection of radio emission if we assume a hollow cone beam centered on the magnetic axis.

With four clear peaks and bridge emission between two different pairs of peaks, the $\gamma$-ray pulse profile of PSR J1536$-$4948 is complicated.  The broad radio peaks might suggest a relatively low magnetic inclination angle, but it is difficult to get sharp peaks in most emission models in such geometries.  The second and third peaks show evidence for pulsed $\gamma$-ray emission out to $>$ 25 GeV in Figure \ref{fig:ebandsJ1536}.  \citet{Harding18} have modeled the spectrum and pulse profile of the Vela pulsar out to 100 TeV. In their model, the GeV emission is curvature radiation and the highest energy photons are produced via inverse Compton interactions between accelerated particles and infrared-optical photons.  \citet{Harding18} use the maximum Lorentz factor of accelerated particles necessary to argue against models in which GeV $\gamma$-rays are the result of synchrotron radiation.  While PSR J1536$-$4948 is not as bright as the Vela pulsar, matching the energy-dependent morphology of the pulse profile could serve as another test of these models.

\section{Summary}
\label{sec:Summary}
We report the GMRT discoveries of three MSPs, PSR J0248$+$4230, PSR J1207$-$5050 and PSR J1536$-$4948, at the positions of unassociated \emph{Fermi}-LAT sources, 4FGL J0248.6+4230, 4FGL J1207.4-5050 and 4FGL J1536.4-4948 respectively. Considering the discovery of four MSPs with $\gamma$-ray associations, \cite [one MSP, J1544$+$4937, published in][]{bh13}, the millisecond pulsar per square degree discovery rate for the \emph{Fermi} directed targeted survey with the GMRT is $\sim$ 0.01. However, the discovery rate is more if we count three more MSPs, that are serendipitously discovered during this survey (discussed in Section \ref{sec:sources}), as well as discoveries by PSC that were among these 375 sources and confirmed by the GMRT. The GMRT interferometric array was successfully used to localise these MSPs in the image plane with $\sim$two orders of magnitude better accuracy than the discovery position associated with the \emph{Fermi} error boxes. These precise positions allowed us to conduct sensitive, follow-up timing observations in phased array mode at 322 and 607 MHz while optimising telescope time usage. PSR J0248$+$4230 and PSR J1536$-$4948 were detected both at 322 and 607 MHz observing frequencies with spectral indices of $-$3.34(6) and $-$2.86(6) respectively. Spectral index for these two MSPs are steeper than general MSP population reported by \cite{kramer98,dai15}. In this context, \cite{frail16} explored steep spectrum radio sources as possible pulsar candidates and discovered new radio MSPs associated with Fermi-LAT sources. However, PSR J1207$-$5050 was only detected at 607 MHz, indicating that the radio spectrum of this MSP possibly turns over between 322$-$607 MHz. A detailed spectral study of this MSP with the upgraded GMRT wide band system is in progress.  

We have presented phase-connected timing models for each MSP from $\sim$ 5 years of radio observations, as well as $\sim$ 11.6 years of $\gamma$-ray TOAs for PSR J1536$-$4948. PSR J0248$+$4230 and PSR J1207$-$5050 are isolated MSPs, whereas PSR J1536$-$4948 is in a binary system with orbital period of $\sim$ 62 days and companion mass of $\sim$ 0.32 M$_\sun$ for an inclination of 60\degr~. We report a weak (2$\sigma$) detection of the orthometric amplitude of the Shapiro delay ($h_3$), which is not enough to determine the mass of the pulsar or mass of the companion in absence of other post-Keplerian parameters. Ongoing coherently dedispersed observations of these MSPs using the ugpraded GMRT will allow us to reduce the TOA uncertainities and will enable better constrains on the binary parameters. This may lead to possible determination of Shapiro delay ($h_3$) for PSR J1536$-$4948 with a higher significance and in turn allow us to determine pulsar and companion masses of the binary system.

In this paper, we also reported the discovery of $\gamma$-ray pulsations from these three MSPs, which confirms that the pulsars are the engines powering the previously unassociated $\gamma$-ray sources. For some of the relatively weak $\gamma$-ray sources associated radio pulsars are relatively bright, indicating that radio flux is uncorrelated with the $\gamma$-ray flux and even faint new LAT sources can harbor bright radio MSPs. Such detections provide strong justification to continue radio observations as new unassociated LAT sources are revealed in analysis of longer data sets. 

Ongoing radio polarimetric studies of these MSPs will be helpful to probe the possible emission geometry enabling further constraints on possible models explaining the observed radio and $\gamma$-ray emission. Profile modeling will also be aided by ongoing investigation of profile evolution of these MSPs for wider radio frequency range with the upgraded GMRT. The ongoing timing observations with the upgraded GMRT will reveal the prospect of using these MSPs in the pulsar timing array which will be reported in a future publication. To conclude, in this paper we present the discovery of three radio MSPs with the GMRT in \emph{Fermi} directed targeted searches. The discovery was followed by long term radio timing and subsequent discovery of $\gamma$-ray pulsations. We also present a study of phase aligned radio, $\gamma$-ray profiles of these MSPs. In addition, we provide a list of target pointings and the detection limits for the \emph{Fermi}-LAT point sources that were observed with the GMRT, which will help to plan future observations for these sources.
\newpage
\renewcommand{\thetable}{A-\arabic{table}}
\setcounter{table}{0}


\begin{appendix}
\label{sec:appendix}

We conducted \textit{Fermi} directed searches with the GMRT during between 2010 November and 2013 September as part of an effort coordinated by the \textit{Fermi} PSC. 
Based on several criteria such as the $\gamma$-ray spectral index, the amount of variability seen, the significance of detection etc, the PSC has rank-ordered the unassociated $\gamma$-ray sources according to the probability of them being pulsars. Out of these we considered sources with $|b| > 5 \degr$ to limit the effects of scatter broadening. In addition we choose a greater fraction of the sources in the declination range $-$40\degr~to $-$53\degr, which is outside the sky coverage of other active PSC searches (e.g. GBT, Effelsberg). Since the LAT point source catalogs evolved during the span of 2011$-$2013, we used the most updated source lists (which sometimes were internal source lists that were not published) for the GMRT observations. During 2010$-$2011 we 
targeted sources in 1FGL catalog that was based on 11 months of LAT data \citep{abdo10}. Whereas during 2012 we used 2FGL catalog  \cite{nolan12}, based on 2FGL catalog. During 2013 we used a three year internal source list prepared by the LAT team and chose comparatively weaker $\gamma$-ray sources and many promising high and mid Galactic latitude sources were still left to be searched for millisecond pulsations. 

Table \ref{src_detail} present details of the GMRT observations for all 375 \emph{Fermi}-LAT sources in this survey. This table includes the observing epoch, frequency and duration.  Additionally, to guide planning of future follow up observations, we have included a 10$\sigma$ detection limit for each source, calculated using the radiometer equation \citep{lorimer04} with the GMRT ETC calculator\footnote{http://www.ncra.tifr.res.in/etc}.


\vspace{0.3cm}
\label{tbl:aux}

\begin{longtable}{|l|c|c|c|c|c|c|c|c|c|c|c|c|}
\caption{Summary of the GMRT observations \\$\dagger$: 10$\sigma$ detection threshold calculated with the GMRT ETC\label{src_detail}}\\
\hline 
\multicolumn{1}{|c|}{\textbf{Source}}	& \multicolumn{1}{|c|}{\textbf{Right ascension}}	&
\multicolumn{1}{|c|}{\textbf{Declination}}	& \multicolumn{1}{|c|}{\textbf{MJD}}	&
\multicolumn{1}{|c|}{\textbf{Frequency}}	& \multicolumn{1}{|c|}{\textbf{Duration}}	&
\multicolumn{1}{|c|}{\textbf{S$_{min}$$^\dagger$}} \\
\multicolumn{1}{|c|}{\textbf{ }}	& \multicolumn{1}{|c|}{\textbf{ }}	& \multicolumn{1}{|c|}{\textbf{ }}	&	\multicolumn{1}{|c|}{\textbf{ }}	& \multicolumn{1}{|c|}{\textbf{(MHz)}}	&
\multicolumn{1}{|c|}{\textbf{(min)}}	& \multicolumn{1}{|c|}{\textbf{(mJy)}} \\
\hline
\endfirsthead

\hline
\multicolumn{1}{|c|}{\textbf{Source}} & \multicolumn{1}{|c|}{\textbf{Right ascension}} &
\multicolumn{1}{|c|}{\textbf{Declination}} & \multicolumn{1}{|c|}{\textbf{MJD}} &
\multicolumn{1}{|c|}{\textbf{Frequency}} & \multicolumn{1}{|c|}{\textbf{Duration}} &
\multicolumn{1}{|c|}{\textbf{S$_{min}$$^\dagger$}} \\
\multicolumn{1}{|c|}{\textbf{ }} & \multicolumn{1}{|c|}{\textbf{ }} & \multicolumn{1}{|c|}{\textbf{ }} &
\multicolumn{1}{|c|}{\textbf{ }} & \multicolumn{1}{|c|}{\textbf{(MHz)}} &
\multicolumn{1}{|c|}{\textbf{(min)}} & \multicolumn{1}{|c|}{\textbf{(mJy)}} \\
\hline
\endhead

\hline \multicolumn{7}{|r|}{{Continued on next page}} \\
\hline
\endfoot

\hline
\endlastfoot
       
J2323-4919	& 23$^\mathrm{h}$23$^\mathrm{m}$02$^\mathrm{s}$.3	& $-$49\degr19\arcmin18\arcsec	& 55521.5969518940	& 607	& 25	& 0.6 \\
J2356.0-5253	& 23$^\mathrm{h}$56$^\mathrm{m}$01$^\mathrm{s}$2	& $-$52\degr53\arcmin59\arcsec	& 55521.6151301238	& 607	& 29	& 0.5 \\
J2330.3-4745	& 23$^\mathrm{h}$30$^\mathrm{m}$23$^\mathrm{s}$.28	& $-$47\degr45\arcmin45\arcsec	& 55521.6361103883	& 607	& 31	& 0.5 \\
18M4913	& 23$^\mathrm{h}$28$^\mathrm{m}$10$^\mathrm{s}$	& $-$40\degr35\arcmin2\arcsec	& 55521.6905664423	& 607	& 30	& 0.5 \\
J0001.9-4158	& 00$^\mathrm{h}$01$^\mathrm{m}$55$^\mathrm{s}$.8	& $-$41\degr58\arcmin54\arcsec	& 55521.7144827111 & 607 & 31  & 0.5 \\
18M4597	& 21$^\mathrm{h}$33$^\mathrm{m}$51$^\mathrm{s}$.8	& $+$66\degr46\arcmin57\arcsec	& 55521.7461526166	& 607	& 34 	& 0.5 \\
18M4955	& 23$^\mathrm{h}$45$^\mathrm{m}$48$^\mathrm{s}$	& $-$15\degr51\arcmin29\arcsec	& 55521.7756292463	& 607	& 15	& 0.8 \\
J2339.7-0531	& 23$^\mathrm{h}$39$^\mathrm{m}$42$^\mathrm{s}$.8	& $-$05\degr31\arcmin25\arcsec	& 55521.7864004526	& 607	& 26 	& 0.6 \\
\hline
\end{longtable}
\vspace{1cm}



\end{appendix}

\acknowledgments
We acknowledge support of the Department of Atomic Energy, Government of India, under project no.12-R\&D-TFR-5.02-0700. The GMRT is run by the National Centre for Radio Astrophysics of the Tata Institute of Fundamental Research, India. We acknowledge the support of GMRT telescope operators for observations. 
 We also acknowledge the generous support of the HPC systems of IUCAA and NCRA. 
 The \emph{Fermi}-LAT Collaboration acknowledges generous ongoing support from a number of agencies and institutes that have supported both the development and the operation of the LAT as well as scientific data analysis.  These include the National Aeronautics and Space Administration and the Department of Energy in the United States, the Commissariat \`a l'Energie Atomique and the Centre National de la Recherche Scientifique / Institut National de Physique Nucl\'eaire et de Physique des Particules in France, the Agenzia Spaziale Italiana and the Istituto Nazionale di Fisica Nucleare in Italy, the Ministry of Education, Culture, Sports, Science and Technology (MEXT), High Energy Accelerator Research Organization (KEK) and Japan Aerospace Exploration Agency (JAXA) in Japan, and the K.~A.~Wallenberg Foundation, the Swedish Research Council and the Swedish National Space Board in Sweden. This work performed in part under DOE Contract DE-AC02-76SF00515.

Additional support for science analysis during the operations phase is gratefully acknowledged from the Istituto Nazionale di Astrofisica in Italy and the Centre National d'\'Etudes Spatiales in France.
The National Radio Astronomy Observatory is a facility of the National Science Foundation operated
under cooperative agreement by Associated Universities, Inc.
PCCF gratefully acknowledges continuing support from the Max Planck Society.
SMR is a CIFAR Fellow and is supported
by the NSF Physics Frontiers Center award 1430284.
We thank Ismael Cognard, Philippe Bruel and Gu\"{o}laugur J\'ohannesson for their comments.
BB acknowledges the comments from Dale Frail and Dave Thompson.

\textit{Fermi}-LAT work at NRL is supported by NASA.

\facilities{GMRT, Fermi LAT}

\software{\sc{GMRT Software Backend}\citep{roy10}, \sc{presto}\citep{ransom02}, \sc{Tempo}\citep{Nice15}, \sc{Tempo2}\citep{Edwards06}, \sc{Dracula}\citep{fr18}}


\end{document}